\documentclass[rnote]{aa}     

\usepackage{graphicx}
\usepackage{txfonts}
\usepackage{natbib}
\usepackage{url}
\bibpunct{(}{)}{;}{a}{}{,}

\newcommand{\be}{\begin{equation}}
\newcommand{\ee}{\end{equation}}

\def\apjl{ApJL}
\def\apjs{ApJS}

\newcommand{\Mpc}{$h^{-1}$\thinspace Mpc}

\newcommand{\vmh}{h^{-1}\mathrm{Mpc} }

\begin{document}  

\title{Characteristic density contrasts in the evolution of superclusters. 
The case of A2142 supercluster
}

\author {Mirt~Gramann\inst{1} 
\and Maret~Einasto\inst{1} 
\and Pekka~Hein\"am\"aki\inst{2}
\and Pekka~Teerikorpi\inst{2}
\and Enn~Saar\inst{1,3}
\and Pasi~Nurmi\inst{2}
\and Jaan~Einasto\inst{1,3,4} 
}

\institute{Tartu Observatory, Observatooriumi 1, 61602 T\~oravere, Estonia
\and 
Tuorla Observatory, University of Turku, V\"ais\"al\"antie 20, Piikki\"o, Finland
\and
Estonian Academy of Sciences, Kohtu 6, 10130 Tallinn, Estonia
\and
ICRANet, Piazza della Repubblica 10, 65122 Pescara, Italy
}

\authorrunning{Gramann et al. }

\offprints{M. Gramann}

\date{ Received   / Accepted   }

\titlerunning{Density contrasts}

\abstract
{
The formation and evolution 
of the cosmic web in which galaxy superclusters
are the largest relatively isolated objects  is governed 
by a gravitational attraction of  dark matter and antigravity  
of dark energy (cosmological constant). 
}
{
We study the characteristic density
contrasts in the spherical collapse model
for several epochs in the supercluster evolution and their dynamical state. 
}
{
We analysed the density contrasts for the turnaround, 
 future collapse, and  zero gravity in different 
$\Lambda$CDM models and applied them to study the dynamical state of the
supercluster A2142 with an almost spherical main body, 
making it a suitable test object to apply  a model that assumes sphericity.}
{
We present characteristic density contrasts in the spherical collapse model
for different cosmological parameters. The analysis of the supercluster
A2142 shows that its high-density core has already 
started to collapse. The zero-gravity
line outlines the outer region of the main body of the supercluster.
In the course of future evolution, the supercluster
may split into several collapsing systems. 
}
{ 
The various density contrasts presented in our study 
and applied to the supercluster A2142 offer a promising way 
to characterise the dynamical 
state and expected future evolution of galaxy superclusters.
}


\maketitle

\section{Introduction} 
\label{sect:intro} 

One  of  the  most  remarkable achievements  of  contemporary  cosmology  is the
discovery of the cosmic web, which is   
a complex hierarchical network of galaxy systems
in which galaxies, galaxy groups, clusters, and superclusters form interconnected
systems separated by voids of various sizes
\citep{1977TarOP...1A...1J, 1978IAUS...79..241J, 1982Natur.300..407Z}.
In the cosmic web, the largest relatively isolated systems are galaxy superclusters
\citep{1956VA......2.1584D, 1958ApJS....3..211A, 1978MNRAS.185..357J, 
1993ApJ...407..470Z, 1994MNRAS.269..301E}. 

The evolution and dynamical state of superclusters 
have been analysed with several methods.  
One of them is the spherical collapse model. 
This model describes the evolution of a 
spherically symmetric perturbation in an expanding universe. 
Under the assumption of sphericity, 
the dynamics of a collapsing shell is determined by the mass in its interior. 
The spherical collapse model has been discussed in detail by
\citet{1934PNAS...20..169T, 1947MNRAS.107..410B, 1972ApJ...176....1G,
1980lssu.book.....P}.
This model has been used in the Press-Schechter 
formalism to evaluate 
the mass function of clusters \citep{1974ApJ...187..425P} and to analyse 
the infall of nearby galaxies into the Virgo cluster 
\citep[e.g.][]{2014ApJ...782....4K}. 
A  spherical collapse model has been applied to study, for example, 
the Corona Borealis supercluster \citep{1998ApJ...492...45S, 2014MNRAS.441.1601P}, 
the Shapley supercluster \citep{2000AJ....120..523R, 2006A&A...447..133P},
and the A2199 supercluster \citep{2002AJ....124.1266R}.

One essential moment in the evolution of a spherical perturbation 
is called turnaround, the moment when the sphere stops expanding together with the universe 
and the collapse begins. 
At the turnaround, the perturbation decouples 
entirely from the Hubble flow of the homogeneous background. 
\citet{2002MNRAS.337.1417G} studied the dynamical state of superclusters 
in different $\Lambda$CDM models 
and showed that only a small 
fraction of superclusters or their high-density cores 
have already reached the turnaround radius and have started 
to collapse at the present epoch.
Unlike clusters, superclusters have not reached a 
quasi-equilibrium configuration.

In the universe with a critical mass density all  
regions with overdensity would eventually collapse. 
In the standard $\Lambda$CDM model, the universe makes the transition 
from a matter-dominated to a dark-energy dominated stage at 
the redshift $z \approx 0.3$. 
As long as the matter density was 
substantially larger than the dark energy density, 
it dominated the evolution of the universe, decelerating 
the expansion and driving the formation of structures with 
gravitational instability. When the average matter density fell below 
two times the dark energy density (at the redshift $z \approx 0.7$),
the dark energy started accelerating  the expansion,
and the formation of structure slowed down. 
At the present epoch, when the acceleration of the 
expansion has recently started, the largest bound structures 
are just forming. In the future evolution of the universe, 
these bound systems separate from each 
other at an accelerating rate, forming isolated ‘‘island universes’’
\citep{2005MNRAS.363L..11B, 2006MNRAS.366..803D, 2009MNRAS.399...97A}.
\citet{2006MNRAS.366..803D} analysed the spherical collapse model 
to find the minimum mass density required for a spherical shell 
to remain bound until a very distant future in an accelerating universe. 
This density criterion can be used to identify the superclusters 
that will eventually turnaround and collapse in the future. 
\citet{2011MNRAS.415..964L} used this approach to construct 
catalogues of superclusters from the SDSS data, and 
\citet{2013MNRAS.429.3272C, 2014A&A...567A.144C}  
applied a similar concept to define X-ray superclusters. 
The mass density required for 
the future collapse is smaller than the density required 
for the turnaround 
and collapse at  the present epoch. 

The presence of the dark energy along with  gravitating matter 
influences the formation of the large-scale structure at all 
scales from groups of galaxies to superclusters. 
\citet{2001PhyU...44.1099C} 
introduced a zero-gravity scale, which describes the relation between 
the gravity force and the antigravity force due to  dark energy. 
At the zero-gravity distance, gravity is equal to antigravity. 
This defines the 
minimum mass density for a gravitationally bound system at  the present epoch
\cite[see also ][and references therein]{2015A&A...577A.144T}.

The aim of this {\it Research Note} is to analyse the 
different characteristic densities in the evolution of superclusters
applying the spherical collapse model. 
We study the densities for the turnaround, 
future collapse, and for zero gravity  in different 
$\Lambda$CDM models (Sect.~\ref{sect:methods}). 
We apply our results to the galaxy supercluster A2142 
\citep[][hereafter E2015]{2015A&A...580A..69E} in Sect.~\ref{sect:scl}. 
The A2142 supercluster has a close to spherical high-density main body. 
This makes it a suitable object to apply methods, which assume sphericity
for the  study of its dynamical state.
 
We use the Hubble parameter $H_0=100~h$ km~s$^{-1}$ Mpc$^{-1}$.

\section{Characteristic density contrasts}
\label{sect:methods} 

The density perturbation in the volume $V$ can be calculated as
\begin{equation}
\Delta\rho = \rho/\rho_{\mathrm{m}},
\end{equation} 
where  $\rho = M/V$ is the matter density in the volume and 
$\rho_{\mathrm{m}} = \Omega_{\mathrm{m}}\rho_{\mathrm{crit}} =
 3 \Omega_{\mathrm{m}} H_0^2 / 8\pi G$ 
is the mean matter density in the local universe. 
For a spherical volume $V = 4\pi R^3/3 $ one can find that
\begin{equation}
\Delta\rho = 0.86\times 10^{-12}\Omega_{\mathrm{m}}^{-1}~(\frac{M}{h^{-1}M_\odot})~(\frac{R}{h^{-1}~{\mathrm{Mpc}}})^{-3}.
\label{eq:sph}
\end{equation}

From Eq.~(\ref{eq:sph}) we can estimate the mass of a structure as
\begin{equation}
M(R) = 1.16\times 10^{12}~\Omega_{\mathrm{m}}\Delta\rho~
(R/h^{-1}~{\mathrm{Mpc}})^{3}h^{-1}M_\odot. 
\label{eq:mass1}
\end{equation} 

Table~\ref{tab:denper} summarises the characteristic density contrasts
in different $\Lambda$CDM models for
$\Omega_{\mathrm{m}}$ values $\Omega_{\mathrm{m}} = 0.3, 0.27$, and $1.0$.
The value $0.3$ is suggested by the Planck results
\citep[$\Omega_{\mathrm{m}} = 0.308 \pm 0.012$, ][]{2015arXiv150201589P},  
$\Omega_{\mathrm{m}} = 0.27$ was
used in E2015 in the study of the A2142 supercluster
(below), and  $\Omega_{\mathrm{m}} = 1$ is used for heuristic reasons
\citep[see also ][]{2002sgd..book.....M}. 
Table~\ref{tab:denper} shows  the density contrast $\rho$/$\rho_{\mathrm{m}}$ 
and also the density contrasts $\rho$/$\rho_{\mathrm{\Lambda}}$, where 
$\rho_{\mathrm{\Lambda}} = \Lambda/8~\pi~G = \Omega_{\mathrm{\Lambda}}~\rho_{\mathrm{crit}}$ 
is the density in the local universe, which corresponds to the
cosmological constant (dark energy), and 
$\rho$/$\rho_{\mathrm{crit}}$. We consider the flat cosmological
models, where $\Omega_{\mathrm{\Lambda}} = 1 - \Omega_{\mathrm{m}}$.

\subsection{Turnaround}
\label{sect:sph}

The spherically averaged radial velocity around a system 
in the shell of radius $R$ can be written as $u = HR - v_{\mathrm{pec}}$, 
where $v_{\mathrm{H}} = HR$ is the Hubble expansion velocity and 
$v_{\mathrm{pec}}$ is 
the averaged radial peculiar velocity towards the centre of the system. 
At the turnaround point, the peculiar velocity $v_{\mathrm{pec}} = HR$ and $u = 0$. 
If $v_{\mathrm{pec}} < HR$, the system expands, 
and if $v_{\mathrm{pec}} > HR$, the system begins to collapse. 

In the spherical collapse model, the peculiar velocity $v_{\mathrm{pec}}$ 
is directly related 
to the density contrast $\Delta\rho$.  
If $\Delta\rho > \Delta\rho_{\mathrm{T}}$, then 
the perturbed region ceases to expand and begins to collapse. 
For $\Omega_{\mathrm{m}} = 1$, 
the density fluctuation at the turnaround point is 
$\Delta\rho_{\mathrm{T}} = (3\pi/4)^{2} = 5.55$ 
\citep[see also ][]{2002sgd..book.....M}. 
The values of the turnaround parameters for various cosmological models 
were given in \citet{2015A&A...575L..14C}. 

For $\Omega_{\mathrm{m}} = 0.27$ and $\Omega_{\mathrm{\Lambda}} = 0.73$  
the density perturbation at the turnaround point is 
$\Delta\rho_{\mathrm{T}} =  13.1$. In this case,   
the mass of a structure  at the turnaround point is
\begin{equation}
M_{\mathrm{T}}(R) = 4.1\times 10^{12}~(R/h^{-1}~{\mathrm{Mpc}})^{3}h^{-1}M_\odot.
\label{eq:mrt}
\end{equation} 

The mass $M_{\mathrm{T}}(R)$ 
describes the minimum mass needed in the sphere with radius $R$ 
for the turnaround and collapse \citep{2002MNRAS.337.1417G}.
For the same value of $\Omega_{\mathrm{m}}$, 
the turnaround density in the open model is
somewhat higher than in the flat model 
\citep[see e.g.][]{1989AJ.....98..755R,
2002MNRAS.337.1417G}. However, the effect of the dark energy on
$\Delta\rho_{\mathrm{T}}$ is small.

\subsection{Future collapse}
\label{sect:fvs}

\citet{2006MNRAS.366..803D} studied the spherical collapse 
in the flat $\Lambda$CDM model
with $\Omega_{\mathrm{m}} = 0.3$
($\Delta\rho_{\mathrm{T}} = 12.2$, see Table~\ref{tab:denper}). 
They found that the minimum density 
required for a shell to remain bound in this model 
is $\rho  = 7.86~\rho_{\mathrm{m}}$ 
($2.36$ times the critical density $\rho_{\mathrm{crit}}$). 
All systems where the matter density is 
$\rho = (7.86 - 12.2)~\rho_{\mathrm{m}}$ 
continue to expand at a decelerating rate and eventually 
turnaround and collapse in the future. 
At the present epoch, the minimum density for a shell to remain 
bound is $\rho = 7.86~\rho_{\mathrm{m}} = 3.37~\rho_{\mathrm{\Lambda}}$. 
At the final state, 
when this shell will turnaround and collapse,
the density in the shell is $\rho_{\mathrm{f}}  = 2~\rho_{\mathrm{\Lambda}}$
\citep{2006MNRAS.366..803D}.

The density criterion $\rho  > 7.86~\rho_{\mathrm{m}}$ 
was used by \citet{2009MNRAS.399...97A} to identify the superclusters 
in the numerical simulations. 
They analysed the future evolution of these kinds of superclusters 
from the present time to an expansion factor $a = 100$. 
The spatial distribution of the superclusters remained essentially 
the same after the present epoch, reflecting the halting growth of 
the cosmic web as the dark energy starts to dominate. 
During  evolution, the superclusters become more spherical 
and clusters in superclusters may merge into one cluster.

Density contrasts for the future collapse for different cosmological 
models were studied by \citet{2015A&A...575L..14C}, 
who showed that for $\Omega_{\mathrm{m}} = 0.27$
the density contrast $\Delta\rho_{\mathrm{FC}} =  8.73$. 
In this case, the minimum mass of the structure that will 
turnaround and collapse in the future is
\begin{equation}
M_{\mathrm{FC}}(R) = 2.7\times 10^{12}~(R/h^{-1}~{\mathrm{Mpc}})^{3}h^{-1}M_\odot.
\label{eq:mrfvs}
\end{equation} 

The future evolution of superclusters depends on the properties 
of the dark energy. For the same value of $\Omega_{\mathrm{m}}$, 
the density contrast for the future collapse in the open model 
can be substantially smaller than that in the flat model 
\citep[see e.g.][]{2015A&A...575L..14C}. 
In these models the growth of the structure also slows down 
when the curvature starts to dominate, 
but this process is not as rapid as in the models with the dark energy. 
In the open model with $\Omega_{\mathrm{m}} = 0.3$, 
the $\Delta\rho_{\mathrm{FC}} =  2.87$ only. 

\subsection{Zero gravity}
\label{sect:zg}

We can write the force affecting a test particle with mass $m$ 
as the sum of Newton's gravity force
produced by the mass $M$ and 
Einstein's antigravity force due to the dark energy \citep{2009A&A...507.1271C}, 
\be F(R) = \big (- \frac{GM}{R^2} + \frac{8\pi G}{3}
\rho_{\Lambda} R \big)~m 
= \frac{4~\pi}{3}~GR~(-\rho + 2~\rho_{\Lambda})~m.
\label{eq:fr} \ee

The gravity and antigravity are equal if $\rho = 2~\rho_{\Lambda}$
(density contrast $\rho/\rho_{m} = 2~\Omega_{\Lambda}/\Omega_{\mathrm{m}}$).
In this case, the acceleration around the system in the shell of
radius $R$ is $du/dt = 0$. If $\rho < 2 \rho_{\Lambda}$, the system expands
with an accelerating rate ($du/dt > 0$).
This criterion can be used to define   the minimum mass density 
for a gravitationally bound system at the present epoch 
 \citep{2012A&A...539A...4C, 2015A&A...577A.144T}. 
 For the flat model with $\Omega_{\mathrm{m}} = 0.27$, 
 we find that the density contrast 
 $\Delta\rho_{\mathrm{ZG}} =  5.41$. 
In this case, the mass of the structure is 
\begin{equation}
M_{\mathrm{ZG}}(R) = 1.7\times 10^{12}~(R/h^{-1}~{\mathrm{Mpc}})^{3}h^{-1}M_\odot.
\label{eq:mrzg}
\end{equation}

If the mass of the system is smaller than $M_{\mathrm{ZG}}(R)$ then 
the system is not gravitationally bound.

In the standard $\Lambda$CDM model, 
the mass density required for the future collapse is higher 
than the mass density for a gravitationally bound system at the 
present epoch. Not all gravitationally bound systems 
remain bound in the future. During  evolution systems 
with $\rho = (5.41 - 8.73)~\rho_{\mathrm{m}}$  continue 
to expand along with the universe. At the final state, when the 
shells with $\rho_{\mathrm{f}} = 2~\rho_{\mathrm{\Lambda}}$  turnaround, 
the zero-gravity scale coincides with the turnaround scale.
 
Table~\ref{tab:denper} presents the characteristic density contrasts for the turnaround, 
 future collapse, and  zero gravity. 
For comparison, we also calculated  the characteristic density contrasts 
for the virialised systems. 
The virial densities were calculated using the approximation derived 
by \citet{1998ApJ...495...80B}. 

We also show the characteristic density contrasts 
for the linear mass scale  in Table~\ref{tab:denper}. 
In this case, the matter density is equal to the mean matter 
density and $\Delta\rho = 1$. The radial velocity around a system 
reaches  the Hubble velocity  $u = HR$ ($v_{\mathrm{pec}} = 0$). 
This linear scale is also called  the 
Einstein-Straus radius 
\citep{1945RvMP...17..120E, 2015A&A...577A.144T}. 
From Eq.~(\ref{eq:mass1}) we find that the linear mass of the structure, which  
corresponds to the mean density for $\Omega_{\mathrm{m}} = 0.27,$ is
\begin{equation}
M_{\mathrm{L}}(R) = 0.3\times 10^{12}~(R/h^{-1}~{\mathrm{Mpc}})^{3}h^{-1}M_\odot.
\label{eq:mrla}
\end{equation} 

For the underdense regions, the mass of the structure 
$M(R) < M_{\mathrm{L}}(R)$.

\begin{table}[ht]
\caption{Characteristic density contrasts.}
\begin{tabular}{rrrr} 
\hline\hline  
(1)&(2)&(3)&(4)\\      
\hline 
 $\Omega_{\mathrm{m}}$ & $\rho$/$\rho_{\mathrm{m}}$ &
 $\rho$/$\rho_{\mathrm{\Lambda}}$ & $\rho$/$\rho_{\mathrm{crit}}$ \\
\hline
Virial         &        &       &       \\  
 1.0           & 178    & -     & 178   \\  
 0.3           & 340    & 146   & 102   \\  
 0.27          & 360    & 133   & 97    \\
 \hline                          
Turn-around    &        &       &       \\  
 1.0           & 5.55   & -     & 5.55  \\  
 0.3           & 12.2   & 5.21  & 3.65   \\  
 0.27          & 13.1   & 4.85  & 3.54  \\
\hline                          
Future collapse&        &       &       \\  
 0.3           & 7.86   & 3.37  & 2.36   \\  
 0.27          & 8.73   & 3.23  & 2.36  \\
\hline                          
Zero gravity   &        &       &       \\  
 0.3           & 4.67   & 2.0   & 1.40  \\  
 0.27          & 5.41   & 2.0   & 1.46   \\
\hline                          
Linear         &        &       &       \\  
 0.3           & 1.0    &  0.43 & 0.3   \\  
 0.27          & 1.0    &  0.37 & 0.27  \\
\hline
\label{tab:denper}  
\end{tabular}\\
\tablefoot{
See Sect.~\ref{sect:methods} for column explanations.
}
\end{table}

\section{Application to the A2142 supercluster}
\label{sect:scl} 

We apply the methods described above to study the dynamical
state of the A2142 supercluster. It is the supercluster
SCl~001 at the redshift $z \approx 0.09$ 
from \citet{2012A&A...539A..80L} supercluster catalogue in which
superclusters of galaxies were determined on the
basis of the luminosity density field of the Sloan Digital Sky Survey (SDSS)
MAIN sample galaxies. \citet{2012A&A...539A..80L} showed that in the supercluster
SCl~001 the luminosity
density (calculated with $8$~\Mpc\ smoothing length, $D8$) 
is the highest in the whole SDSS MAIN survey region.
Recently, E2015
presented a detailed study of this supercluster
proposing to call it the A2142 supercluster, according 
to its richest galaxy cluster \object{A2142}.
Details about the supercluster can be found in E2015.
In accordance with E2015 we use  the following standard cosmological parameters below: 
the matter density $\Omega_{\rm m} = 0.27$ and the 
dark energy density $\Omega_{\Lambda} = 0.73$ \citep{2011ApJS..192...18K}.

The morphology of this and other superclusters
from \citet{2012A&A...539A..80L} catalogue was studied in 
\citet{2011A&A...532A...5E} who showed that 
the A2142 supercluster consists of a
quite spherical main body with outgoing straight filament-like tail. 
This makes the A2142 supercluster, and especially its main body,
a suitable test object to apply methods that assume sphericity.

\begin{figure}[ht]
\centering
\resizebox{0.44\textwidth}{!}{\includegraphics[angle=0]{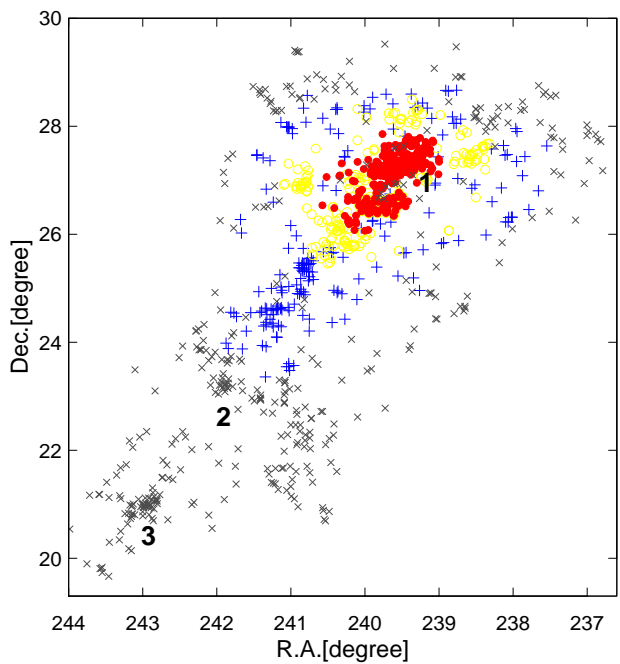}}
\caption{
Distribution of galaxies in the A2142 supercluster in the sky plane 
in global density regions as described in the text. 
Red filled circles denote galaxies in the region of global
density $D8 \geq 17$; yellow empty circles correspond to galaxies with global density
$13 \leq D8 < 17$. Blue crosses correspond to galaxies with global density
$8 \leq D8 < 13$, and grey Xs galaxies with $5 \leq D8 < 8$.
The size of the highest density region is approximately $1.8$
degrees, and the size of the region
with $D8 \geq 13$ is approximately $3$~degrees; sizes in megaparsecs are given in 
Table~\ref{tab:D8prop}.
Number 1 marks the Abell cluster A2142, and numbers 2 and 3  indicate two regions of
galaxy groups in the tail of the supercluster, as explained in the text.
}
\label{fig:radecd13}
\end{figure}

E2015 analysed the luminosity density distribution
in the A2142 supercluster and showed that it can be divided into 
four global density regions. Two high-density regions form a high-density
core of the supercluster, lower density regions form the outskirts of the 
supercluster. Figure~\ref{fig:radecd13} presents the sky distribution of galaxies 
in these global luminosity density regions, and
Table~\ref{tab:D8prop} summarises their properties. 

The A2142 supercluster with its outgoing filament is quite asymmetrical,
therefore we apply a spherical collapse model to the main body of the supercluster,
and to the two group regions in the tail of the supercluster,
as explained below.
In addition to regions defined on the basis of the luminosity density,
we choose the main body of the supercluster as follows:
we use luminosity
density limit $D8 = 5$ \citep[this is the luminosity density limit used to define
superclusters in ][]{2012A&A...539A..80L}, and exclude two regions
with galaxy groups from the tail of the superclusters, denoted as (2) and (3)
in Fig.~\ref{fig:radecd13}. These two regions are analysed separately
(see E2015 for details about these regions). 
In addition, we study 
the local environment of the main body of the supercluster denoted as
Main+env in Table~\ref{tab:D8prop}. 
This region has no strict density limit.
The radii of the regions are given in Table~\ref{tab:D8prop}.
Table~\ref{tab:D8prop} shows that, in addition to the main body of the supercluster, 
the "Main+env" region 
only includes the close neighbourhood of its main body,
approximately $2$~\Mpc\ from the supercluster boundaries.
This is underdense region penetrated by filaments of galaxy groups
and single galaxies outgoing from the supercluster. 
These outgoing filaments were shortly discussed in E2015. 

Masses of different global density regions are calculated as follows.
Data about galaxy groups in the A2142 supercluster were taken from the group
catalogue by \citet{2014A&A...566A...1T}, who also calculated  group
masses on the basis of the virial theorem. We summed group masses in each region,
which gives the dynamical mass of the region. We also calculated the estimated masses. 
In each region there are some single galaxies. They may be the brightest galaxies
of faint groups in which other member galaxies are too faint to be 
observed within SDSS survey magnitude limits.
We used the median mass of groups
with less than five member galaxies in the supercluster
as the mass of these faint groups. 
To obtain the total mass of these faint groups, this median
mass was multiplied with the number of single galaxies in a region. 
In addition, we added 10\% of the total mass 
to the mass of the region as the mass of  intracluster gas. 
The details and references of the calculations of dynamical and estimated masses
can be found in E2015. The supercluster's total estimated
mass is approximately 1.5 times larger than its dynamical mass. 
This agrees quite well with the estimates given in \citet{2014A&A...567A.144C},
who compared supercluster masses from observations and simulations. 
In Table~\ref{tab:D8prop}, we give the density contrasts 
of these global density regions calculated 
using dynamical and estimated masses and radii of these regions.

\begin{table}[ht]
\setlength{\tabcolsep}{4pt}
\caption{Density contrasts, masses,  and radii of the core and 
the outskirts regions of the A2142 supercluster.}
\begin{tabular}{rrrrrr} 
\hline\hline  
(1)&(2)&(3)&(4)&(5)&(6)\\      
\hline 
 $D8$ &   $M_{\mathrm{dyn}}$&  $M_{\mathrm{est}}$ & $R$ 
 & $\Delta\rho_{\mathrm{dyn}}$ & $\Delta\rho_{\mathrm{est}}$ \\
     & [$10^{15}h^{-1}M_\odot$] & [$10^{15}h^{-1}M_\odot$]& &&\\ 
\hline
Full scl         & 2.9 & 4.3 &       &      \\  
 $ \geq 17$      & 1.2 & 1.4 & 6     & 17.9 & 20.9\\
 $ \geq 13$      & 1.7 & 2.1 & 8     & 10.5 & 13.0\\
 $ \geq 8 $      & 2.2 & 3.0 &10     &  7.1 &  9.7 \\
 Main ($ \geq 5 $) & 2.5 & 3.7 & 13  &  3.6 &  5.3 \\
 Main+env          & 2.8 & 4.6 & 15  &  2.6 &  3.2 \\
\hline                                        
 (2)             & 0.10 & 0.18 & 2.5 & 20.4 & 36.7 \\
 (3)             & 0.30 & 0.42 & 2.5 & 61.2 & 85.7 \\
\hline
\label{tab:D8prop}  
\end{tabular}\\
\tablefoot{
Columns are as follows:
(1): Global luminosity density $D8$ of the regions (in units of mean luminosity 
density, 
$\ell_{\mathrm{mean}}$ = 1.65$\cdot10^{-2}$ $\frac{10^{10} h^{-2} L_\odot}{(\vmh)^3}$). 
Main body of the supercluster is defined as $D8 \geq 5$, without group regions
2 and 3. Main+env denotes main body of the supercluster without group regions 2 and 3,
but with local neighbourhood of the main body; see text for details);
(2): the total dynamical mass of groups (in case of groups with 2 and 3 member galaxies
we use median mass); 
(3): the total estimated mass of a region (including faint groups and intracluster
    gas, see text); 
(4): radius of the region (in $h^{-1}$ Mpc);
(5): density contrast of the region according to the dynamical mass;
(6): density contrast of the region according to the  estimated mass.   
}
\end{table}

\begin{figure}[ht]
\centering
\resizebox{0.440\textwidth}{!}{\includegraphics[angle=0]{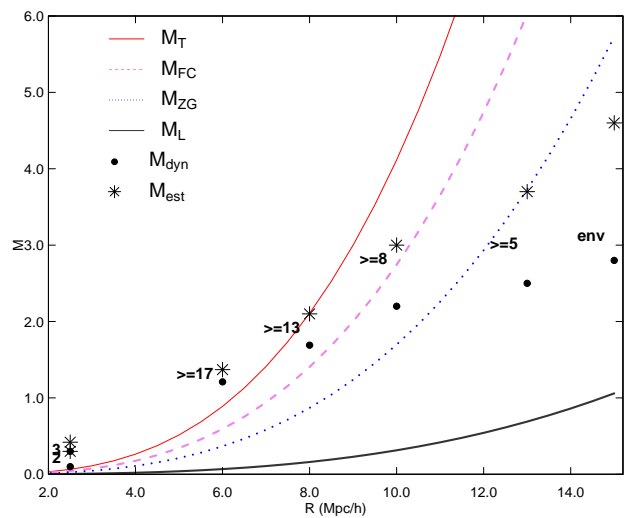}}
\caption{
Mass corresponding to the turnaround mass $M_{\mathrm{T}}(R)$ (red line),
future collapse mass $M_{\mathrm{FC}}(R)$ (violet line), 
zero-gravity mass $M_{\mathrm{ZG}}(R)$ (blue line), and linear mass
$M_{\mathrm{L}}(R)$ (grey line;
 in units of $10^{15}h^{-1}M_\odot$) 
versus radius of a sphere $R$  in different dynamical evolution models
for $\Omega_{\mathrm{m}} = 0.27$.
Filled circles show the total masses of galaxy groups in regions of different global 
density in the A2142 supercluster (Table~\ref{tab:D8prop}).   
Stars denote estimated masses as explained in the text.
Numbers show global density lower limit for a region 
($env$ marks Main+env region, 2 and 3 denote regions of galaxy groups in the tail
of the supercluster).
}
\label{fig:mt}
\end{figure}

In Fig.~\ref{fig:mt} we plot  mass $M(R)$ 
versus radius of a sphere $R$ for different characteristic
density contrasts
(Eq.~(\ref{eq:mrt}),~(\ref{eq:mrfvs}),~(\ref{eq:mrzg}), and~(\ref{eq:mrla}))
and show both the dynamical and estimated
masses of global density regions in the A2142 supercluster
(in Fig.~\ref{fig:mt} Main+env region is denoted as "env").

Figure~\ref{fig:mt} shows that points corresponding to the highest density
core of the A2142 supercluster ($D8 \geq 17$) lie well above
the turnaround line, suggesting that this region has started to collapse.
The density contrasts of this region are high (Table~\ref{tab:D8prop}).
The cluster A2142 lies in this region.
We note that most of the mass in this region comes from the mass
of the cluster A2142 with dynamical mass 
$M_{dyn} = 0.9\times~10^{15}h^{-1}M_\odot$ in \citet{2014A&A...566A...1T}. 
This value is almost equal to the mass
found by \citet{2014A&A...566A..68M} using several methods
(see also  discussion in E2015).
Two group regions from the tail of the supercluster
may have started to collapse, too.  

According to the estimated mass, 
the full high-density core of the supercluster with $D8 \geq 13$
may have reached the turnaround or is close to it. Lower global density 
regions do not have high enough mass for the present-day collapse.
However, the estimated mass of the region with $D8 \geq 8$ 
is high enough for this region to start
the collapse in the future, as the FC line in Fig.~\ref{fig:mt}
and the density contrast in Table~\ref{tab:D8prop} show.
The outer parts 
of the supercluster with $5 \leq D8 \leq 8$ may separate from
the main body in the future.
The filamentary tail of the supercluster
(including group regions 2 and 3)
lies below global density level $D8 = 8$ and probably does not
collapse in the future  with supercluster main body.
If we add this structure to the main body 
of the supercluster then the mass, as well as the volume, of the
system increases. The increase in 
mass is not large enough for this system to
collapse as a whole at present or in the future (Fig.~\ref{fig:mt}). 
Taking into account  that regions 2 and 3 collapse 
themselves, the most likely scenario 
is that these regions (or at least the region 3) will separate from 
supercluster's main body.
Thus Fig.~\ref{fig:mt} suggests that the A2142 supercluster
may split into several systems in the course of the future
evolution.

According to the estimated mass, the gravity-antigravity balance line
(zero gravity) in Fig.~\ref{fig:mt}
borders the main body of the supercluster. Points corresponding
to the region, which also includes
the close environment of the supercluster lie below the zero gravity
line suggesting that this region is not gravitationally bound.
The linear mass scale lies farther away from the supercluster
and its close neighbourhood. 
The zero-gravity line in Fig.~\ref{fig:mt} 
presents in another way some of the information
given by the $\Lambda$ significance graph 
($\log \rho/\rho_{\Lambda}$ vs. $\log R$)
introduced by \citet{2015A&A...577A.144T}.  
This curve corresponds
to the horizontal straight line $\rho/ \rho_{\Lambda} = 2$ 
in the $\Lambda$ significance graph. 
The points lying above this line correspond to the high-density regions
of the superclusters, as shown in the present paper for the A2142 supercluster.

\section{Summary}
\label{sect:summary} 

We used the spherical collapse model to study the
characteristic density contrasts in the evolution of superclusters.
Despite its idealised nature, the spherical collapse model has been
popular because of its simplicity. Also, this model gives a reasonable 
description of the collapse of high peaks in a random Gaussian density
field \citep[see e.g.][and references therein]
{1994ApJ...427...51B, 2010gfe..book.....M}.

We analysed the density contrasts for the turnaround, 
future collapse, and  zero gravity in the $\Lambda$CDM models.
The turnaround marks the epoch when the peculiar velocity
reaches the Hubble velocity ($v_{\mathrm{pec}} = HR$) and the radial velocity
around the system is $u = 0$. The mass density for the future
collapse shows the density required for the perturbed region
to continue to expand with an decelerating rate and eventually
turnaround and collapse in the future.
The zero gravity marks the epoch
when the radial acceleration around the system is $du/dt = 0$:
after that the perturbed region starts to expand with an
accelerating rate. 

Typically, rich superclusters have a complicated structure and the models which 
assume sphericity can be applied to their central parts only 
\citep[see e.g.][for an analysis of supercluster morphology]{2011A&A...532A...5E}.
The collapsing high-density cores have been studied 
in  the very rich  Shapley and Corona Borealis superclusters
\citep{2000AJ....120..523R, 2014MNRAS.441.1601P}, 
in the Perseus-Pisces supercluster \citep{2001A&A...378..345H, 2015A&A...577A.144T},
in the A2199 supercluster, the member
of the Hercules supercluster \citep{2002AJ....124.1266R, 2001AJ....122.2222E},
and also in the richest supercluster of the Sloan Great Wall
\citep[][and Hein\"am\"aki et al., in preparation]{2007A&A...476..697E, 
2011ApJ...736...51E}.
\citet{2015A&A...575L..14C} analysed the Shapley supercluster 
with the future collapse model and concluded that only the central part of the
Shapley supercluster will form a supercluster in the distant future.

We applied our results to 
study the dynamical state of the supercluster A2142 with an almost
spherical main body.
Our analysis showed that the high-density core of the A2142 supercluster has already 
started to collapse. This conclusion was reached also by E2015.
In the course of future evolution the supercluster
may split into several collapsing systems. 
\citet{2009MNRAS.399...97A} showed that during the future evolution,
groups and clusters in superclusters may merge. In the A2142 supercluster,
merging groups have been observed in the cluster A2142 region
\citep{2014A&A...570A.119E} and perhaps in the outskirts of the
high-density core of the supercluster \citep{2011ApJ...727L..38K}.
In the A2142 supercluster,
the zero-gravity line borders the main body of the supercluster.
This supports the conclusion that the  
A2142 supercluster does not remain bound in the future, its outer parts and tail
may separate from the main body of the supercluster. 
Also, the close neighbourhood of the supercluster is not bound to the supercluster
and may continue to expand together with the universe.

Density contrasts analysed in our study can be used to study the dynamical
state and future evolution of a large sample of galaxy superclusters.
The study of the
galaxy properties in dynamically different regions may
provide interesting insight for the environmental studies of the galaxies
in superclusters. 
We will continue the study of the dynamical state of galaxy superclusters
from observations and simulations in a forthcoming paper.

\section*{Acknowledgments}

We thank the referee for comments and suggestions 
that helped to improve the paper.
We are pleased to thank the SDSS Team for the publicly available data
releases.  Funding for the Sloan Digital Sky Survey (SDSS) and SDSS-II has been
provided by the Alfred P. Sloan Foundation, the Participating Institutions,
the National Science Foundation, the U.S.  Department of Energy, the
National Aeronautics and Space Administration, the Japanese Monbukagakusho,
and the Max Planck Society, and the Higher Education Funding Council for
England.  The SDSS Web site is \texttt{http://www.sdss.org/}.
The SDSS is managed by the Astrophysical Research Consortium (ARC) for the
Participating Institutions.  The Participating Institutions are the American
Museum of Natural History, Astrophysical Institute Potsdam, University of
Basel, University of Cambridge, Case Western Reserve University, The
University of Chicago, Drexel University, Fermilab, the Institute for
Advanced Study, the Japan Participation Group,  Johns Hopkins University,
the Joint Institute for Nuclear Astrophysics, the Kavli Institute for
Particle Astrophysics and Cosmology, the Korean Scientist Group, the Chinese
Academy of Sciences (LAMOST), Los Alamos National Laboratory, the
Max-Planck-Institute for Astronomy (MPIA), the Max-Planck-Institute for
Astrophysics (MPA), New Mexico State University, Ohio State University,
University of Pittsburgh, University of Portsmouth, Princeton University,
the United States Naval Observatory, and the University of Washington.

The present study was supported by the ETAG projects 
IUT26-2 and IUT40-2, and by the European Structural Funds
grant for the Centre of Excellence "Dark Matter in (Astro)particle Physics and
Cosmology" TK120. This work has also been supported by
ICRAnet through a professorship for Jaan Einasto.

\bibliographystyle{aa}
\bibliography{mirt.bib}

\begin{thebibliography}{50}
\expandafter\ifx\csname natexlab\endcsname\relax\def\natexlab#1{#1}\fi

\bibitem[{{Abell}(1958)}]{1958ApJS....3..211A}
{Abell}, G.~O. 1958, \apjs, 3, 211

\bibitem[{{Araya-Melo} {et~al.}(2009){Araya-Melo}, {Reisenegger}, {Meza}, {van
  de Weygaert}, {D{\"u}nner}, \& {Quintana}}]{2009MNRAS.399...97A}
{Araya-Melo}, P.~A., {Reisenegger}, A., {Meza}, A., {et~al.} 2009, \mnras, 399,
  97

\bibitem[{{Bernardeau}(1994)}]{1994ApJ...427...51B}
{Bernardeau}, F. 1994, \apj, 427, 51

\bibitem[{{Bondi}(1947)}]{1947MNRAS.107..410B}
{Bondi}, H. 1947, \mnras, 107, 410

\bibitem[{{Bryan} \& {Norman}(1998)}]{1998ApJ...495...80B}
{Bryan}, G.~L. \& {Norman}, M.~L. 1998, \apj, 495, 80

\bibitem[{{Busha} {et~al.}(2005){Busha}, {Evrard}, {Adams}, \&
  {Wechsler}}]{2005MNRAS.363L..11B}
{Busha}, M.~T., {Evrard}, A.~E., {Adams}, F.~C., \& {Wechsler}, R.~H. 2005,
  \mnras, 363, L11

\bibitem[{{Chernin}(2001)}]{2001PhyU...44.1099C}
{Chernin}, A.~D. 2001, Physics Uspekhi, 44, 1099

\bibitem[{{Chernin} {et~al.}(2009){Chernin}, {Teerikorpi}, {Valtonen},
  {Dolgachev}, {Domozhilova}, \& {Byrd}}]{2009A&A...507.1271C}
{Chernin}, A.~D., {Teerikorpi}, P., {Valtonen}, M.~J., {et~al.} 2009, \aap,
  507, 1271

\bibitem[{{Chernin} {et~al.}(2012){Chernin}, {Teerikorpi}, {Valtonen},
  {Dolgachev}, {Domozhilova}, \& {Byrd}}]{2012A&A...539A...4C}
{Chernin}, A.~D., {Teerikorpi}, P., {Valtonen}, M.~J., {et~al.} 2012, \aap,
  539, A4

\bibitem[{{Chon} {et~al.}(2014){Chon}, {B{\"o}hringer}, {Collins}, \&
  {Krause}}]{2014A&A...567A.144C}
{Chon}, G., {B{\"o}hringer}, H., {Collins}, C.~A., \& {Krause}, M. 2014, \aap,
  567, A144

\bibitem[{{Chon} {et~al.}(2013){Chon}, {B{\"o}hringer}, \&
  {Nowak}}]{2013MNRAS.429.3272C}
{Chon}, G., {B{\"o}hringer}, H., \& {Nowak}, N. 2013, \mnras, 429, 3272

\bibitem[{{Chon} {et~al.}(2015){Chon}, {B{\"o}hringer}, \&
  {Zaroubi}}]{2015A&A...575L..14C}
{Chon}, G., {B{\"o}hringer}, H., \& {Zaroubi}, S. 2015, \aap, 575, L14

\bibitem[{{de Vaucouleurs}(1956)}]{1956VA......2.1584D}
{de Vaucouleurs}, G. 1956, Vistas in Astronomy, 2, 1584

\bibitem[{{D{\"u}nner} {et~al.}(2006){D{\"u}nner}, {Araya}, {Meza}, \&
  {Reisenegger}}]{2006MNRAS.366..803D}
{D{\"u}nner}, R., {Araya}, P.~A., {Meza}, A., \& {Reisenegger}, A. 2006,
  \mnras, 366, 803

\bibitem[{{Eckert} {et~al.}(2014){Eckert}, {Molendi}, {Owers}, {Gaspari},
  {Venturi}, {Rudnick}, {Ettori}, {Paltani}, {Gastaldello}, \&
  {Rossetti}}]{2014A&A...570A.119E}
{Eckert}, D., {Molendi}, S., {Owers}, M., {et~al.} 2014, \aap, 570, A119

\bibitem[{{Einasto} {et~al.}(1994){Einasto}, {Einasto}, {Tago}, {Dalton}, \&
  {Andernach}}]{1994MNRAS.269..301E}
{Einasto}, M., {Einasto}, J., {Tago}, E., {Dalton}, G.~B., \& {Andernach}, H.
  1994, \mnras, 269, 301

\bibitem[{{Einasto} {et~al.}(2001){Einasto}, {Einasto}, {Tago}, {M{\"u}ller},
  \& {Andernach}}]{2001AJ....122.2222E}
{Einasto}, M., {Einasto}, J., {Tago}, E., {M{\"u}ller}, V., \& {Andernach}, H.
  2001, \aj, 122, 2222

\bibitem[{{Einasto} {et~al.}(2015){Einasto}, {Gramann}, {Saar},
  {Liivam{\"a}gi}, {Tempel}, {Nevalainen}, {Hein{\"a}m{\"a}ki}, {Park}, \&
  {Einasto}}]{2015A&A...580A..69E}
{Einasto}, M., {Gramann}, M., {Saar}, E., {et~al.} 2015, \aap, 580, A69

\bibitem[{{Einasto} {et~al.}(2011{\natexlab{a}}){Einasto}, {Liivam{\"a}gi},
  {Tago}, {Saar}, {Tempel}, {Einasto}, {Mart{\'{\i}}nez}, \&
  {Hein{\"a}m{\"a}ki}}]{2011A&A...532A...5E}
{Einasto}, M., {Liivam{\"a}gi}, L.~J., {Tago}, E., {et~al.} 2011{\natexlab{a}},
  \aap, 532, A5

\bibitem[{{Einasto} {et~al.}(2011{\natexlab{b}}){Einasto}, {Liivam{\"a}gi},
  {Tempel}, {Saar}, {Tago}, {Einasto}, {Enkvist}, {Einasto}, {Mart{\'{\i}}nez},
  {Hein{\"a}m{\"a}ki}, \& {Nurmi}}]{2011ApJ...736...51E}
{Einasto}, M., {Liivam{\"a}gi}, L.~J., {Tempel}, E., {et~al.}
  2011{\natexlab{b}}, \apj, 736, 51

\bibitem[{{Einasto} {et~al.}(2007){Einasto}, {Saar}, {Liivam{\"a}gi},
  {Einasto}, {Tago}, {Mart{\'{\i}}nez}, {Starck}, {M{\"u}ller},
  {Hein{\"a}m{\"a}ki}, {Nurmi}, {Gramann}, \&
  {H{\"u}tsi}}]{2007A&A...476..697E}
{Einasto}, M., {Saar}, E., {Liivam{\"a}gi}, L.~J., {et~al.} 2007, \aap, 476,
  697

\bibitem[{{Einstein} \& {Straus}(1945)}]{1945RvMP...17..120E}
{Einstein}, A. \& {Straus}, E.~G. 1945, Reviews of Modern Physics, 17, 120

\bibitem[{{Gramann} \& {Suhhonenko}(2002)}]{2002MNRAS.337.1417G}
{Gramann}, M. \& {Suhhonenko}, I. 2002, \mnras, 337, 1417

\bibitem[{{Gunn} \& {Gott}(1972)}]{1972ApJ...176....1G}
{Gunn}, J.~E. \& {Gott}, III, J.~R. 1972, \apj, 176, 1

\bibitem[{{Hanski} {et~al.}(2001){Hanski}, {Theureau}, {Ekholm}, \&
  {Teerikorpi}}]{2001A&A...378..345H}
{Hanski}, M.~O., {Theureau}, G., {Ekholm}, T., \& {Teerikorpi}, P. 2001, \aap,
  378, 345

\bibitem[{{J{\~o}eveer} \& {Einasto}(1978)}]{1978IAUS...79..241J}
{J{\~o}eveer}, M. \& {Einasto}, J. 1978, in IAU Symposium, Vol.~79, Large Scale
  Structures in the Universe, ed. M.~S. {Longair} \& J.~{Einasto}, 241--250

\bibitem[{{J{\~o}eveer} {et~al.}(1978){J{\~o}eveer}, {Einasto}, \&
  {Tago}}]{1978MNRAS.185..357J}
{J{\~o}eveer}, M., {Einasto}, J., \& {Tago}, E. 1978, \mnras, 185, 357

\bibitem[{{J{\~o}eveer} {et~al.}(1977){J{\~o}eveer}, {Einasto}, \&
  {Tago}}]{1977TarOP...1A...1J}
{J{\~o}eveer}, M., {Einasto}, J., \& {Tago}, M. 1977, Tartu
  Astrof{\"u}{\"u}s.~Obs.~Preprint, Nr.~A-1, 45 p., 1, A1

\bibitem[{{Karachentsev} {et~al.}(2014){Karachentsev}, {Tully}, {Wu}, {Shaya},
  \& {Dolphin}}]{2014ApJ...782....4K}
{Karachentsev}, I.~D., {Tully}, R.~B., {Wu}, P.-F., {Shaya}, E.~J., \&
  {Dolphin}, A.~E. 2014, \apj, 782, 4

\bibitem[{{Kawahara} {et~al.}(2011){Kawahara}, {Yoshitake}, {Nishimichi}, \&
  {Sousbie}}]{2011ApJ...727L..38K}
{Kawahara}, H., {Yoshitake}, H., {Nishimichi}, T., \& {Sousbie}, T. 2011,
  \apjl, 727, L38

\bibitem[{{Komatsu} {et~al.}(2011){Komatsu}, {Smith}, {Dunkley}, {Bennett},
  {Gold}, {Hinshaw}, {Jarosik}, {Larson}, {Nolta}, {Page}, {Spergel},
  {Halpern}, {Hill}, {Kogut}, {Limon}, {Meyer}, {Odegard}, {Tucker}, {Weiland},
  {Wollack}, \& {Wright}}]{2011ApJS..192...18K}
{Komatsu}, E., {Smith}, K.~M., {Dunkley}, J., {et~al.} 2011, \apjs, 192, 18

\bibitem[{{Liivam{\"a}gi} {et~al.}(2012){Liivam{\"a}gi}, {Tempel}, \&
  {Saar}}]{2012A&A...539A..80L}
{Liivam{\"a}gi}, L.~J., {Tempel}, E., \& {Saar}, E. 2012, \aap, 539, A80

\bibitem[{{Luparello} {et~al.}(2011){Luparello}, {Lares}, {Lambas}, \&
  {Padilla}}]{2011MNRAS.415..964L}
{Luparello}, H., {Lares}, M., {Lambas}, D.~G., \& {Padilla}, N. 2011, \mnras,
  415, 964

\bibitem[{{Mart{\'{\i}}nez} \& {Saar}(2002)}]{2002sgd..book.....M}
{Mart{\'{\i}}nez}, V.~J. \& {Saar}, E. 2002, {Statistics of the Galaxy
  Distribution} (Chapman {\&} Hall/CRC, Boca Raton)

\bibitem[{{Mo} {et~al.}(2010){Mo}, {van den Bosch}, \&
  {White}}]{2010gfe..book.....M}
{Mo}, H., {van den Bosch}, F.~C., \& {White}, S. 2010, {Galaxy Formation and
  Evolution} (Cambridge University Press)

\bibitem[{{Munari} {et~al.}(2014){Munari}, {Biviano}, \&
  {Mamon}}]{2014A&A...566A..68M}
{Munari}, E., {Biviano}, A., \& {Mamon}, G.~A. 2014, \aap, 566, A68

\bibitem[{{Pearson} {et~al.}(2014){Pearson}, {Batiste}, \&
  {Batuski}}]{2014MNRAS.441.1601P}
{Pearson}, D.~W., {Batiste}, M., \& {Batuski}, D.~J. 2014, \mnras, 441, 1601

\bibitem[{{Peebles}(1980)}]{1980lssu.book.....P}
{Peebles}, P.~J.~E. 1980, {The large-scale structure of the universe}
  (Princeton University Press)

\bibitem[{{Planck Collaboration} {et~al.}(2015){Planck Collaboration}, {Ade},
  {Aghanim}, {Arnaud}, {Ashdown}, {Aumont}, {Baccigalupi}, {Banday},
  {Barreiro}, {Bartlett}, \& et~al.}]{2015arXiv150201589P}
{Planck Collaboration}, {Ade}, P.~A.~R., {Aghanim}, N., {et~al.} 2015, ArXiv
  e-print: 1502.01589

\bibitem[{{Press} \& {Schechter}(1974)}]{1974ApJ...187..425P}
{Press}, W.~H. \& {Schechter}, P. 1974, \apj, 187, 425

\bibitem[{{Proust} {et~al.}(2006){Proust}, {Quintana}, {Carrasco},
  {Reisenegger}, {Slezak}, {Muriel}, {D{\"u}nner}, {Sodr{\'e}}, {Drinkwater},
  {Parker}, \& {Ragone}}]{2006A&A...447..133P}
{Proust}, D., {Quintana}, H., {Carrasco}, E.~R., {et~al.} 2006, \aap, 447, 133

\bibitem[{{Regos} \& {Geller}(1989)}]{1989AJ.....98..755R}
{Regos}, E. \& {Geller}, M.~J. 1989, \aj, 98, 755

\bibitem[{{Reisenegger} {et~al.}(2000){Reisenegger}, {Quintana}, {Carrasco}, \&
  {Maze}}]{2000AJ....120..523R}
{Reisenegger}, A., {Quintana}, H., {Carrasco}, E.~R., \& {Maze}, J. 2000, \aj,
  120, 523

\bibitem[{{Rines} {et~al.}(2002){Rines}, {Geller}, {Diaferio}, {Mahdavi},
  {Mohr}, \& {Wegner}}]{2002AJ....124.1266R}
{Rines}, K., {Geller}, M.~J., {Diaferio}, A., {et~al.} 2002, \aj, 124, 1266

\bibitem[{{Small} {et~al.}(1998){Small}, {Ma}, {Sargent}, \&
  {Hamilton}}]{1998ApJ...492...45S}
{Small}, T.~A., {Ma}, C.-P., {Sargent}, W.~L.~W., \& {Hamilton}, D. 1998, \apj,
  492, 45

\bibitem[{{Teerikorpi} {et~al.}(2015){Teerikorpi}, {Hein{\"a}m{\"a}ki},
  {Nurmi}, {Chernin}, {Einasto}, {Valtonen}, \& {Byrd}}]{2015A&A...577A.144T}
{Teerikorpi}, P., {Hein{\"a}m{\"a}ki}, P., {Nurmi}, P., {et~al.} 2015, \aap,
  577, A144

\bibitem[{{Tempel} {et~al.}(2014){Tempel}, {Tamm}, {Gramann}, {Tuvikene},
  {Liivam{\"a}gi}, {Suhhonenko}, {Kipper}, {Einasto}, \&
  {Saar}}]{2014A&A...566A...1T}
{Tempel}, E., {Tamm}, A., {Gramann}, M., {et~al.} 2014, \aap, 566, A1

\bibitem[{{Tolman}(1934)}]{1934PNAS...20..169T}
{Tolman}, R.~C. 1934, Proceedings of the National Academy of Science, 20, 169

\bibitem[{{Zeldovich} {et~al.}(1982){Zeldovich}, {Einasto}, \&
  {Shandarin}}]{1982Natur.300..407Z}
{Zeldovich}, I.~B., {Einasto}, J., \& {Shandarin}, S.~F. 1982, \nat, 300, 407

\bibitem[{{Zucca} {et~al.}(1993){Zucca}, {Zamorani}, {Scaramella}, \&
  {Vettolani}}]{1993ApJ...407..470Z}
{Zucca}, E., {Zamorani}, G., {Scaramella}, R., \& {Vettolani}, G. 1993, \apj,
  407, 470

\end{thebibliography}

\end{document}